\documentclass[pdflatex,sn-mathphys-num]{sn-jnl}

\usepackage{graphicx}%
\usepackage{multirow}%
\usepackage{amsmath,amssymb,amsfonts}%
\usepackage{amsthm}%
\usepackage{mathrsfs}%
\usepackage[title]{appendix}%
\usepackage{xcolor}%
\usepackage{textcomp}%
\usepackage{manyfoot}%
\usepackage{booktabs}%
\usepackage{algorithm}%
\usepackage{algorithmicx}%
\usepackage{algpseudocode}%
\usepackage{listings}%

\theoremstyle{thmstyleone}%
\theoremstyle{thmstyletwo}%

\theoremstyle{thmstylethree}%

\begin{document}
\title[Article Title]{Liquid Droplet as Adaptive Material while Levitating via Coupling between Plasma and Kelvin Force
}


\author[1]{\fnm{Ping-Rui} \sur{Tsai}}
\author[1]{\fnm{Hong-Yue} \sur{Huang}}
\author[2]{\fnm{Ying-Pin} \sur{Tsai}}
\author[3]{\fnm{Chih-Jung} \sur{Lin}}
\author[1]{\fnm{Bo-Kai} \sur{Xu}}
\author[1]{\fnm{Jih-Kang} \sur{Hsieh}}
\author[1]{\fnm{Yu-Ting} \sur{Cheng}}
\author[4]{\fnm{Cheng-Wei} \sur{Lai}}
\author[5]{\fnm{Yu Hsuan} \sur{Kao}}
\author[6]{\fnm{Wen-Chi} \sur{Chen}}
\author[2]{\fnm{Fu-Li} \sur{Hsiao}}
\author[3]{\fnm{Yu-Jane} \sur{Sheng}}
\author[4]{\fnm{Po-Heng} \sur{Lin}}
\author[1*]{\fnm{Tzay-Ming} \sur{Hong}}

\affil[1]{\orgdiv{Department of Physics}, \orgname{National Tsing Hua University}, \orgaddress{\city{Hsinchu}, \postcode{30013}, \state{Taiwan}, \country{R.O.C}}}

\affil[2]{\orgdiv{Institute of Photonics}, \orgname{National Changhua University of Education}, \orgaddress{\city{Changhua}, \postcode{50007}, \state{Taiwan}, \country{R.O.C}}}

\affil[3]{\orgdiv{Department of Chemical Engineering}, \orgname{National Taiwan University}, \orgaddress{\city{Taipei}, \postcode{10617}, \state{Taiwan}, \country{R.O.C}}}

\affil[4]{\orgdiv{Department of Chemistry}, \orgname{National Chung Hsing University}, \orgaddress{ \postcode{402204}, \state{Taiwan}, \country{R.O.C}}}

\affil[5]{\orgdiv{Department of Color and Illumination Technology}, \orgname{National Taiwan University of Science and Technology}, \orgaddress{\city{Taipei}, \postcode{106335}, \state{Taiwan}, \country{R.O.C}}}

\affil[6]{\orgdiv{Institute of Photonics Technologies}, \orgname{National Tsing Hua University}, \orgaddress{ \city{Hsinchu}, \postcode{30013}, \state{Taiwan}, \country{R.O.C}}}

\abstract{  Fascinating in art and science, the ability to float is also captivating and relevant in practical applications, such as Penning and ion traps that are fundamental to quantum computing. In this work, we first reproduce the classic water bridge by glycerol and, as it breaks down due to thermal agitation, observe that a lump of glycerol with mass$\sim$2.5 g can float and exhibit near-periodic oscillations. Through experiments, finite element analysis, and simulations, we discover that the stability of the floating droplet is made possible by the interaction between three mechanisms: Deformation, Plasma, and Kelvin force. 
Note that glycerol cluster (GC) falls in the class of adaptive materials that can change their properties or behavior in response to varying environmental conditions, i.e., stimuli-responsive. Furthermore, the stimuli, modified by the deformation of GC, collaborate with it to create this unique simple, yet stable, floating system.
Backed up by simulations, this process, operated by only a single pair of electrodes, holds the potential to develop a simple yet powerful railgun.}




\maketitle

\section{Introduction}

Floating has always fascinated people. After the success of the Wright flyer in 1903, aviation has become a safe and regular means to travel. Different ways to make levitation possible have been realized in the laboratory, such as the ion trap in quantum computing   \cite{haffner2008quantum} and the Penning trap to contain antimatter  \cite{andresen2007antimatter}, which operates on the principle of generating the lowest potential energy at the trap using multiple sets of electrodes. Optical tweezers employ laser focusing to levitate objects by EM field  \cite{moffitt2008recent}.
Classic demonstrations to counter the force of weight include trapping water droplets and making frogs float by sound wave and magnetic field  \cite{falkovich2005floater,simon2000diamagnetic}. While the former results from the external force, the latter makes use of the diamagnetic response from the body of frog. 

For over 130 years, scientists aim to understand (1) the floating mechanism, (2) the characteristics of internal flow, and (3) the unique structure of WB. One prevailing theory considers surface tension and Kelvin force induced by Maxwell pressure as the main source to resist weight  \cite{woisetschlager2012horizontal,aerov2011water} and maintain the stability of bridge contour\cite{montazeri2013experimental}. Although Raman spectroscopy  \cite{ponterio2010raman} detected differences in spectroscopic signals between WB and water without the external electric field, X-ray analysis  \cite{skinner2012structure} did not reveal any ordered arrangement. However, MRI and PIV experiments identified a distinctive double-layer flow structure within WB  \cite{wexler2017magnetic,tsai2020evidence}. So far, no comprehensive theoretical framework is able to integrate all these experimental findings.

\begin{figure}
\centering
\includegraphics[width=8.5cm]{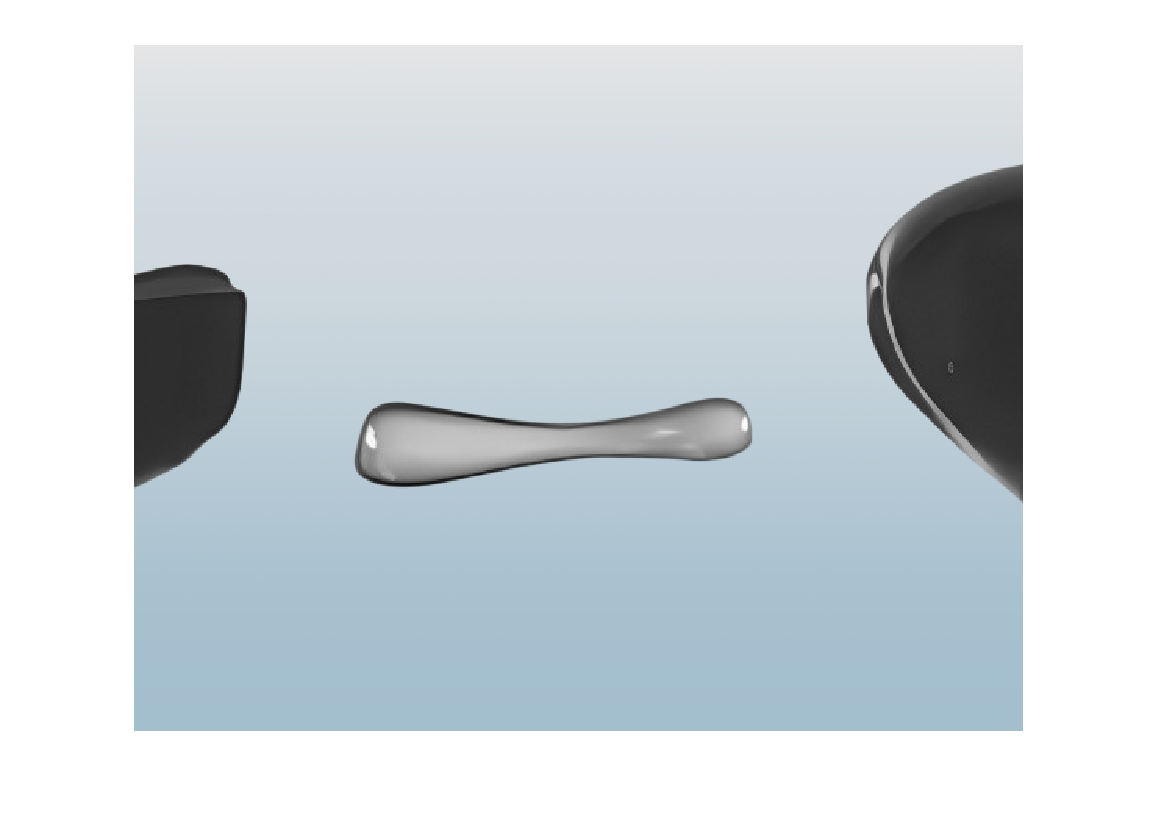}\caption{\textbf{Levitating glycerol clusters.} When glycerol is placed in a setup similar to that of WB, it can generate levitating glycerol clusters in the air without the need for a bridge structure (see Supplementary Information:  Supplementary Video 1 for details).}
\end{figure}  In 1893, the British engineer William Armstrong conducted the elegant water bridge experiment (WB), which is considered a representative example of changing the properties of matter to counteract weight \cite{fuchs2014armstrong}. When we apply a DC voltage of a few kilovolts to a pair of electrodes immersed in two beakers separated by a distance of mm to cm and filled with pure deionized water, a water bridge can automatically form and float across the beakers. This phenomenon also occurs in numerous polar dielectric liquids \cite{woisetschlager2012horizontal}. 
Here, we report that a glycerol cluster (GC) can be made to levitate without the threads or connections to both beakers at room temperature where glycerol ($1.2 $ Pa$\cdot$s) is approximately 10,000 times more viscous than water ($8.9 \times 10^{-4}$ Pa$\cdot$s), as depicted in Fig. 1. Since it shares the same experimental setup as WB, the floating GC offers new insights on the latter phenomenon. 

This new discovery provides a breakthrough in the field of electrohydrodynamics (EHD), with GC simultaneously encompassing two subfields: (1) floating EHD bridges \cite{fuchs2014armstrong}, and (2) how droplets deform or break up in the presence of a strong electric field \cite{o1953distortion,allan1962particle}. The former can be represented by the phenomenon of WB, whereas the latter pertains to the behavior of a single droplet in strong electric fields, which is believed to result from the surface charges induced by molecular polarization. 

For the subfield of deformation theory, Taylor pioneered models for leaky dielectric materials \cite{xu2023breakup}, and subsequent modifications by Saville introduced electrokinetic models suitable for significant droplet deformations \cite{saville1997electrohydrodynamics}. In addition to the effects of polarized charges, the accumulation of ions at both ends under the influence of the electric field counteracts surface tension, resulting in deformation. This phenomenon is famously known as the Taylor cone \cite{fernandez2007fluid}.  Unlike common experiments that involve placing the droplet in oil to keep it afloat \cite{abbasi2020deformation},  GC does not require such a medium.

Our project explores four key topics: (1). How does GC oscillate horizontally and vertically, and what keeps it afloat? (2). How does the deformation of GC trigger and couple with the plasma? (3). How does the Kelvin force–plasma coupling lengthen GC lifetime? (4). Potential application of GC setup in the railgun. The answer to these questions will explain not only the mechanism of levitation, but also the stability of confinement by a mere pair of electrodes, deemed impossible by previous understanding of the electromagnetic trap\cite{andresen2007antimatter}. Section 2.1$\sim$3 involve an intricate coupling between surface tension, Kelvin force, and Plasma also known as ionized gas jet (IGJ)\cite{park2021stabilization}, which together delay the eventual breakup of GC. Section 2.4 builds upon the existing experimental framework of GC and introduces a novel form of electromagnetic railgun for matter\cite{mcnab2005development,xie2021research}, relying solely on voltage and plasma, implemented through a parallel configuration of multiple pairs of electrodes. We demonstrate its feasibility through numerical simulations.


\section{Results}

\begin{figure}
\centering
\includegraphics[width=8.5cm]{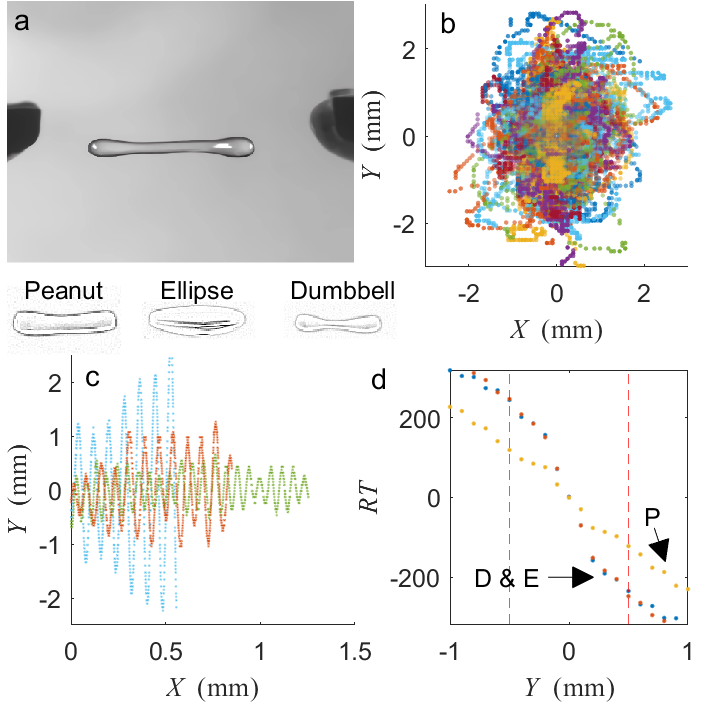}\caption{\textbf{Trajectory of GC and its floating mechanism.} \textbf{a}  GC is shown to float between beakers with  $V$=12000 V and gap distance 0.8 cm. There are roughly three configurations for GC. \textbf{b} The trajectories of the centroid of GC are shown for fourteen samples for each of the six different parameter sets. It is clear that GC is confined in both $X$ and $Y$ directions. \textbf{c} The oscillation  period of $Y$-displacement for $V$= 13000 V varied with different gap distance = 0.6, 0.7, and 0.8 cm, represented by the red, green, and blue lines. \textbf{d} The simulated RT values by COMSOL is shown where P, E, and D refer to the configurations described in \textbf{e}, and the two red vertical dashed lines represent the position of one standard
error of 0.5 mm.}
\end{figure}  

\subsection{How does GC oscillate and what keeps it afloat?}

Data were captured by a high-speed camera  of 3000 fps (Phantom Miro-ex4, Gray-level image) and analyzed by Open CV to track the trajectory of more than 14 sets of GCs in order to find the patterns of their motion under six different parameter sets, labeled as condition 1, 2, and 3  for a gap between beakers of 0.6, 0.7 and 0.8 cm with $V$=12000, and condition 4, 5, and 6 with 13000 V.

 We observed that GC can generally be categorized into three profiles, as shown in Fig. 2a. The trajectories of GC in Fig. 2b show that they are confined and oscillatory in both horizontal and vertical directions during a lifetime of 0.1 $\sim$ 1 sec. After applying a Fast Fourier Transform (FFT) on the vertical displacement in Fig. 2c, we obtained two main peaks, indicative of dual modes of period $\approx 3.3 \times 10^{-4}$ and $6.0\times 10^{-2}$ sec. Based on Fig. 2c where there are approximately 7 oscillation cycles within 0.5 sec, we conclude that  $6.0\times 10^{-2}$ sec is the true period.  
Plugging this period in $M\ddot y= -M(2\pi/T)^2 y$ where $M$ is the mass of GC and  the amplitude  $\approx 0.5$ mm allows us to estimate the magnitude of restoring force and its ratio to $Mg$, denoted by $RT$,  at 1.5. 
The shorter period is likely caused by the intrinsic flow field, which has no preference in direction, because it also appears in FFT for the horizontal or $X$ direction.

Let's first review the origin of Kevin force by defining the polarization density $\vec P$ as
\begin{equation}
\vec P = D \vec p= (\epsilon-\epsilon_{0}) \vec E
 \end{equation}
where $D$ represents the number of dipoles with moment $\vec p$ per unit volume from the polarization constitutive law, and $\epsilon_{0}$ and $\epsilon$ are the vacuum and dielectric permittivity. The Kelvin polarization force density  \cite{montazeri2013experimental,woisetschlager2012horizontal,peng2023review,haus1989electromagnetic}
 \begin{equation}
 {\vec f_{\rm Kelvin}}= \vec P \cdot \nabla \vec E
 \end{equation}
 needs to be calculated before summing over the droplet to give the force experienced by these dipole moments in the presence of a nonuniform electric field.
 Substituting Eq. (1) into (2), we obtain 
  \begin{equation}
\vec f_{\rm Kelvin}=\frac{1}{2}\epsilon_{0}(\epsilon_{r}-1)\nabla(E^{2}).
 \end{equation}

 Based on the finite element method (FEM), the COMSOL Multiphysics (Comsol Inc., Palo Alto, CA) simulates $RT$ from Eq. (3) for the parameters 12000 V and 0.8 cm, which lands $RT$ at roughly $10^2$ for the $Y$ displacement, see Fig. 2d. Since the gap between GC and the sprout of the beaker have reached the electrical breakdown in air at about 3 kV/cm  \cite{tupikin2020combined}, the main drop in electric potential occurs on both sides of GC. Although the restoring force lifting GC was verified, the experimental value for RT differs from the simulation result by two orders of magnitude.


\begin{figure}
\centering
\includegraphics[width=8.5cm]{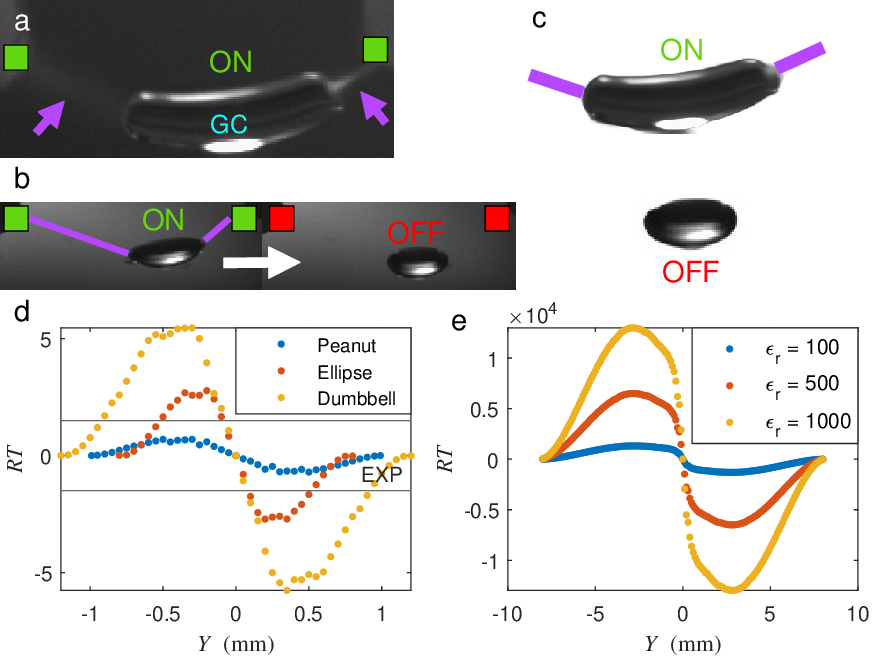}\caption{\textbf{Energy loss due to electrical breakdown and RT correction.} \textbf{a.} The purple arrows indicate the IGJs on both sides of GC, whose presence allows the charges to flow between GC and the beaker spouts whose positions are marked by the green square regions.
\textbf{b.} Note that the deformation of GC is more severe on the left photo when the electrical conduction is enabled by the IGJs highlighted by purple lines. The right photo is obtained when the voltage  drops below the breakdown threshold and the conduction ceases. The square regions at the beaker spout are now painted in red to indicated that it is an open circuit.
\textbf{c.} Snapshots of GC highlight the degree of deformation between \textbf{a} and \textbf{b} (see Supplementary Video 2 $\&$ 3 for details). \textbf{d.} This plot considers RT correction by accounting for the energy loss due to electrical breakdown; the two black lines labeled EXP indicate the experimentally measured RT value. \textbf{e.} RT values obtained under the same experimental conditions as \textbf{d}, but for different relative permittivities.}
\end{figure}  
What causes this discrepancy? Experimentally it is observed that, as soon as the plasma disappears when the spacing between GC and beaker spout widens, GC ceases to deform and begins to drop due to the absence of the Kelvin force. This is captured by Fig.3 a$\sim$c. Therefore, we calculate the product of breakdown voltage and the straight-line distance between the beaker spout and GC. By subtracting the corresponding energy loss from the applied voltage, we can obtain an RT value that matches the experimental results. To provide a reference for materials with different relative permittivities, we conducted simulations by using the same voltage and elliptical contour. This allows us to evaluate how various materials perform in this setup, as detailed by Fig. 3e. 



\begin{figure}
\centering
\includegraphics[width=8.5cm]{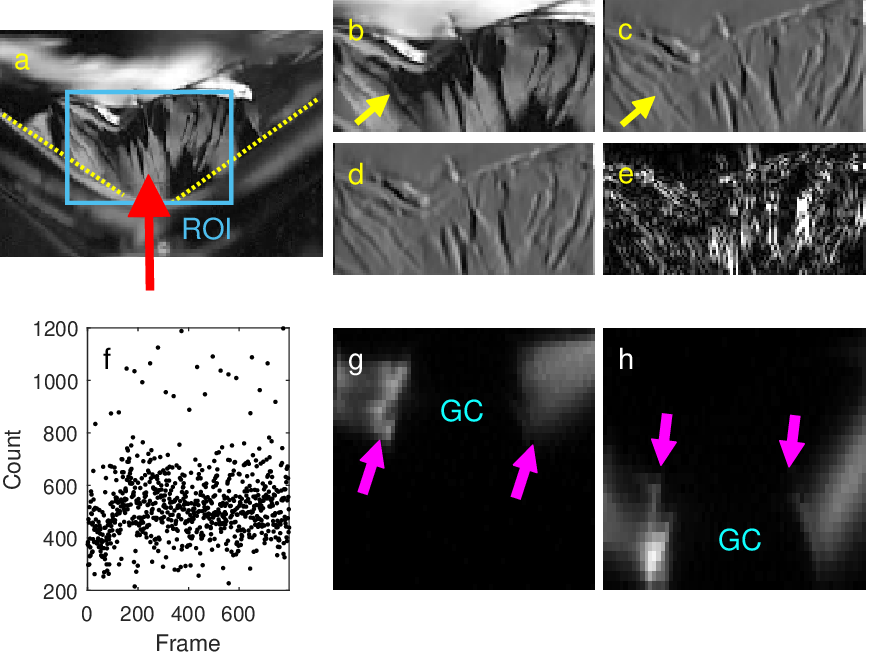}\caption{\textbf{The Impact Pattern of IGJs on the Glycerol Surface.} \textbf{a.} Side view of the glycerol surface that is deformed and pushed up by the plasma bombardment, denoted by the red arrow, near the beaker spout whose edges are marker by yellow dotted lines.
\textbf{b.} The region of interest (ROI) in a. is magnified to show the linear creases riddled on the  plasma-blasted Glycerol surface.
\textbf{c.} shows a clearer view of the contours in b. by employing the Sobel operator; the yellow arrows in both \textbf{b} and \textbf{c} point at the same position.
\textbf{d.} presents another plasma impact contours from a later frame.
\textbf{e.} is the image difference, obtained by subtracting the absolute pixel value of \textbf{c} and \textbf{d}.
\textbf{f.} The total number of pixels on all creases is plotted against the number of video frame. The random distribution is typical of the plasma bombardment.
In \textbf{g.} and \textbf{h.}, our attention is shifted from the beaker to GC where the purple arrows
indicate the IGJs on both sides of GC. They each display the plasma trajectories when GC  returns to its upper and lower positions. As is near the beaker spout, the site of plasma bombardment on GC is selected randomly. See the Supplementary Videos 4 \& 5.}
\end{figure}  

\subsection{How deformation of GC triggers and couples with plasma?}
In Section 2.1, we verified that the floating mechanism arises from the Kelvin force generated by the electric field, which in turn is contributed by the presence of IGJs. the ability of plasma to store and release energy is analogous to the function of a bowstring. However, in addition to providing the lifting force, the electric field also exerts a Coulomb force on the polarized charges on both sides of GC \cite{abbasi2020deformation}, which normally seals its fate of being ruptured. However, this is not what we observed. Therefore, there must be a fuse-like mechanism that protect GC against the rupture.
Since the timing of GC generation is unpredictable and its lifetime is no more than one second, it is challenging to directly track the plasma \cite{kong2012plasma,xian2009plasma,mericam2009experimental}.
The IGJs typically occur within 10–100 ns, and capturing such a fast dynamics requires  specialized intensified high-speed camera with at least $10^7$ fps \cite{kong2012plasma,xian2009plasma,mericam2009experimental,andorICCD}.
Furthermore, to achieve both high spatial and high temporal resolution simultaneously is still technically unfeasible \cite{hao2025multiple,sarasaen2021fine,soylu2021circumventing}. Under these constraints, insights can still be gained by observing the glycerol surface near the beaker spout. The impact creases in Fig. 4(a$\sim$e) clearly show the footprint of plasma discharges, which occur at random position as proved in Fig. 4f. Likewise, the IGJ also left its mark on the GC surface as shown in Fig. 4g and h captured during the oscillation of GC .
\begin{figure}
\centering
\includegraphics[width=8.5cm]{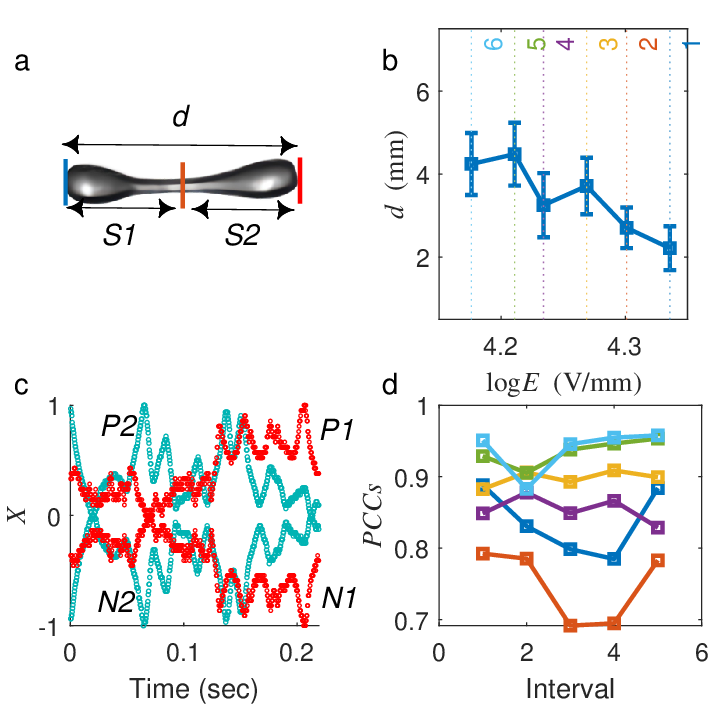}\caption{\textbf{Extension and contraction of GC in the horizontal direction.}  \textbf{a} Length of left (right) arm relative to the centroid is denoted by $S1$ ($S2$), and their sum or the total length of GC by $d$.  \textbf{b} The time-averaged $d$ is plotted against the logarithm of electric field between beakers, with the colored numbers labeling the six conditions of experimental parameters. \textbf{c} The  $P1$ and $N1$ denote the time variation of $S1$ and $S2$ for condition 5, while $P2$ and $N2$ are for condition 6. The maximum amplitude has been normalized to unity to facilitate comparison. \textbf{d} The PCCs is calculated between $S1$ and $S2$ and presented for the six different conditions under parameter sets defined in \textbf{b}.}
\end{figure}  

We are also concerned about the synchronization of plasma impacts on both sides of GC because it is crucial to keep GC from contacting and coalescing with the glycerol in the beakers. For this purpose, we analyzed the extension and length variation along the horizontal major axis of GC in Fig. 5a $\&$ b. The average length, $d$, is found to decrease when we raise the electric field. This is because, within the limits allowed by surface tension, GC  can be further stretched before breakage.  The spacing with the beaker spout is another factor that affects the maximum value of $d$. 

The process of equilibration can be observed in the horizontal oscillation of GC relative to its centroid, as illustrated in Fig. 5c. Since GC  is not a rigid body, the plasma discharge rates on the left and right sides are not necessarily correlated. It turns out that $S1$ and $S2$ in Fig. 5d exhibit a positive correlation across all five intervals, with the Pearson correlation coefficient (PCCs) consistently above 0.7.

To verify the picture regarding the horizontal 
oscillation of GC and its interaction with the plasma\cite{park2021stabilization}, let's start by examining the force equilibrium
\begin{equation}
    F_{\text{plasma}} + F_{\text{Coulomb}} + F_{\text{other}} = 0
\end{equation}
where \( F_{\text{plasma}} \) denotes the force exerted by the IGJ, \( F_{\text{Coulomb}} \) is the electrostatic force due to induced surface polarization charges, and \( F_{\text{other}} \) that includes the vertical gravitational and EHD forces  can be ignored to focus on the horizontal motion.
Roughly, the plasma-induced force can be approximated as:
\begin{equation}
    F_{\text{plasma}} = P_{\text{IGJ}} \cdot A = \sum_{i} n_i m_i v_i^2 A
\end{equation} 
where \( P_{\text{IGJ}} \) is the transient pressure exerted by the IGJ, \( A \) is the effective surface area on which the IGJ impinges, and $n_i$, $m_i$, and $v_i$ denote respectively the number density, mass, and average velocity of particle species $i$ (e.g., ions or electrons) along the direction of propagation. 
In the meantime, the Coulomb force due to surface polarization can be expressed in terms of the local electric field intensity:
\begin{equation}
    F_{\text{Coulomb}} = \frac{1}{2} \varepsilon_0 E^2 A
\end{equation}
where \( \varepsilon_0 \) denotes the vacuum permittivity and \( E \) the vertical electric field near the GC surface.
Assuming that (1) the IGJ exerts momentum flux due to the motion of charged ions and electrons, and (2) this flux is counteracted by the electrostatic pressure arising from surface polarization, the equilibrium condition gives:

\begin{equation}
    \sum_{i} n_i m_i v_i^2 = \frac{1}{2} \varepsilon_0 E^2
\end{equation}


By use of Eq. (7), we employed a hybrid computational approach combining coarse-grained molecular dynamics—specifically, dissipative particle dynamics (DPD)—with Monte Carlo (MC) simulations to verify the proposed mechanism. The former simulates the interactions among charged particles under the influence of an external electric field, while the latter focuses on the effects of plasma on GC. Through trial and error, we found that choosing the probability of plasma bombardment as $P(x) = (x/L)^{2n}$ where $x$ is the particle $X$ coordinate, $L=12$ is a given length scale, and $n$ is set to 50, could best avoid GC from contacting either electrode and reproduce the observed oscillation. Simulation results are shown in Fig. 6a$\sim$e and compare well with the actual observation in Fig. 6f. See Supporting Information Video 6 for details. 


To further investigate how GC evolves from the dumbbell shape to rupture, we designed an additional DPD-MC simulation under the following three initial conditions:

\begin{enumerate}
    \item $L = 12$, The particles impacted by the IGJ will have zero charge and zero horizontal velocity, see Fig. 7a \& d.
    \item $L = 12$, The particles impacted by the IGJ will have zero charge, see Fig. 7b \& e.
    \item $L = 18$, The particles impacted by the IGJ will have zero charge, see Fig. 7c \& f.
\end{enumerate}

In all three cases, the initially charged particles on the left and right sides are equal. Conditions 1 and 2 allow comparison of the effect of IGJ momentum impacting the GC surface on the rupture time. Conditions 2 and 3 allow examination of the influence of plasma hit probability on rupture behavior. From Fig. 7d $\sim$ f, which display the proportion of $\pm2$ charged particles near the center point (or the rupture singularity), it is evident that in both conditions 1 and 2, GC rupture occurs at the center, with condition 1 resulting in a shorter lifetime. In contrast, condition 3 does not exhibit central rupture.

IGJ halts rupture by forming a dumbbell-shaped tear. Acting like ocean waves hitting shallow water, it blocks forward motion. Though front particles lose Coulomb interactions, rear particles—driven by inertia and Lennard-Jones forces—continue stacking, forming clusters on both sides. However, under Condition 3, the Coulomb force acting on the particles far exceeds the blocking momentum imparted by the IGJs, preventing such aggregation. Experimentally, we observe that at the vertical position furthest from the beaker, where the restoring force reaches a maximum, the GC undergoes a pronounced elongation deformation. This is followed by a contraction of the overall length as the system approaches the equilibrium position, as shown in Fig. 7g.



\begin{figure}
\centering
\includegraphics[width=8.5cm]{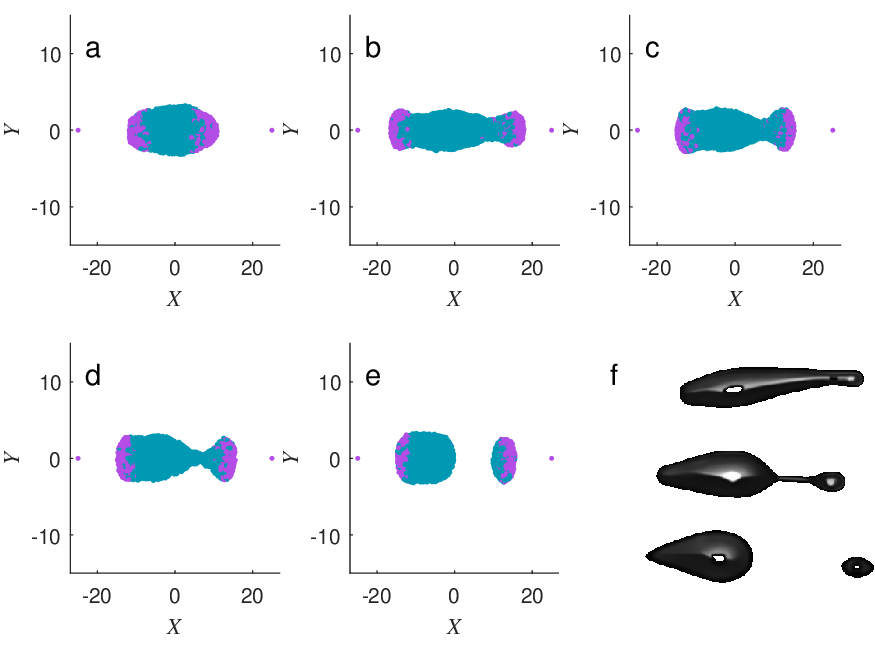}\caption{\textbf{Dissipative Particle Dynamics simulation for a cluster of particles in a strong electric field.} \textbf{a}$\sim$\textbf{e.} show how the cluster deforms and eventually splits over time in the $X$ direction, where purple and light blue colors distinguish charged from neutral ones. 
\textbf{f.} show how GC evolves from the upper to the lower configurations in real experiments.}
\end{figure}

\begin{figure}
\centering
\includegraphics[width=8.5cm]{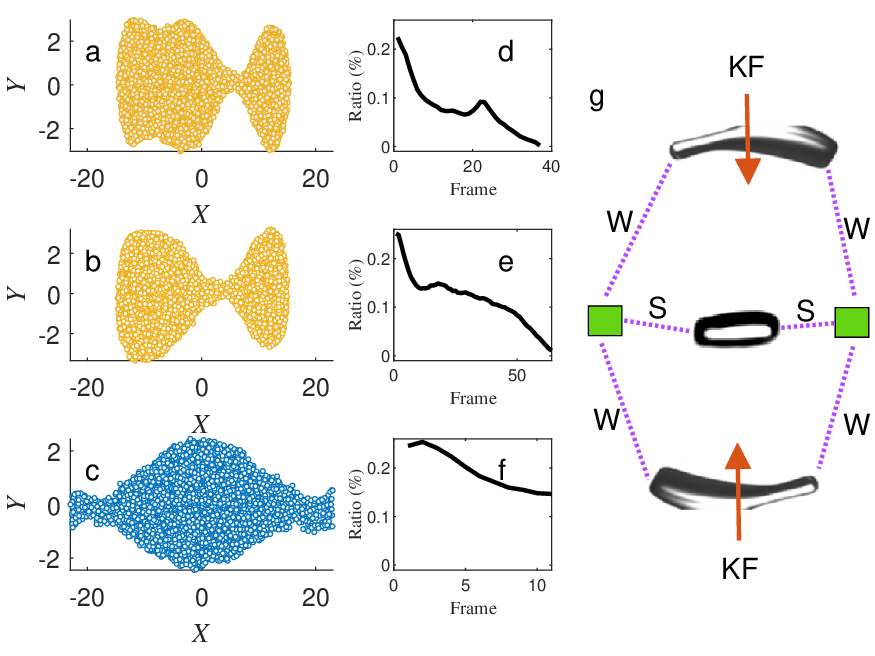}\caption{\textbf{a} $\sim$ \textbf{c} show the deformation differences under three conditions, while \textbf{d} $\sim$ \textbf{f} present the particle distribution ratio at the center and rupture regions for each case. \textbf{g} combines theoretical and experimental insights to illustrate the roles of the Kelvin force (KF), and the zones where the impact frequency of IGJ is weak (W) and strong (S), revealing a positive correlation between horizontal elongation and the center-of-mass motion in the Y-direction.}
\end{figure}  

\subsection{How does the Kelvin force–plasma coupling lengthen GC lifetime?}

\begin{figure}
\centering
\includegraphics[width=8.5cm]{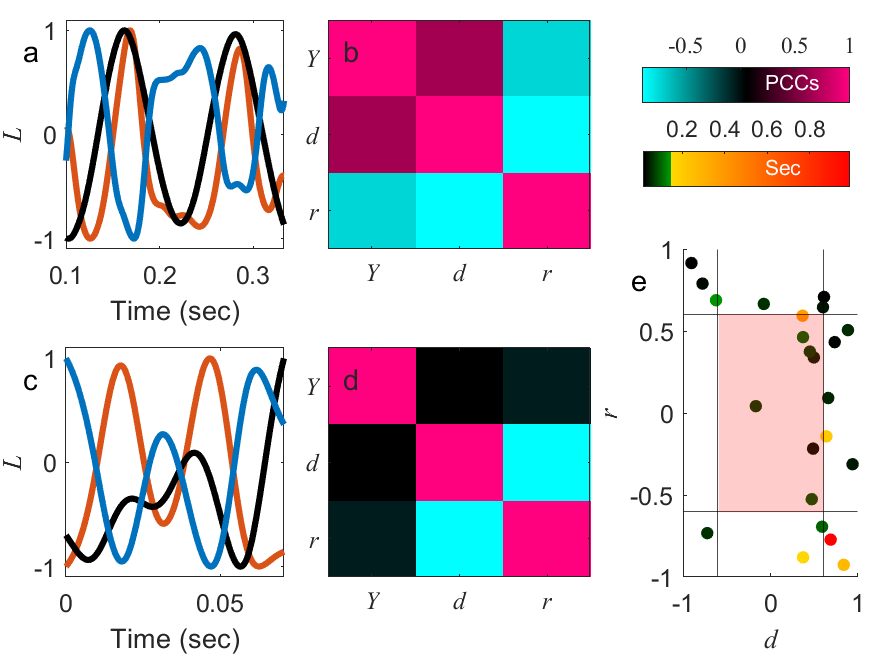}\caption{\textbf{Correlation between vertical displacement ($Y$) and the horizontal length ($d$) and bottleneck radius ($r$) of GC.} These three lengths, denoted by the black, orange and blue lines, are normalized for easy comparison. 
Data with a lifetime close to 1 second are analyzed in \textbf{a} \& \textbf{b}, as opposed to those less than 0.1 second in  \textbf{c} \& \textbf{d}. They show how these lengths evolve with time and their similarity matrices. The latter clearly reveals that the existence of correlations or synchronization among \textbf{$Y$}, $d$, and $r$ are intimately linked to the stability of GC, i.e., plot \textbf{b}. This observation is further verified in Panel \textbf{e} by including 23 samples. Data points inside the light red regime are unstable since they fall within the thresholds of $\pm$0.6.}
\end{figure}  



The simulations in Sec. 2.2 show that the amount of induced dipoles and accumulated ions on both sides of GC depends sensitively on their distance to the beakers. Furthermore, the correlation between the vertical displacement $Y$ and the horizontal length $d$ of GC is expected, but yet unproven. In this section, we explore whether the synchronization among $Y$, $d$, and the bottleneck radius $r$ affects the lifetime of GC by experiments.

The evolution of these three variables and their similarity matrices are plotted in Fig. 8a \& b and c \& d that respectively correspond to lifetime close to 1 and less than 0.1 second, Correlations only exist in a \& b, not c \&d. In other words, the persistence of GC relies on the positive/negative correlation between $Y$ and \textit{d}/\textit{r}.  By including 23 samples, Fig. 8e further shows that the correlation with $d$ is stronger than $c$ since the long-lived GC tend to accumulate at $d>0.6$.
The conclusions from Secs. 2.1 to 2.3 are summarized in Table 1.





\begin{table}[ht]
\centering
\begin{tabular}{|c|c|p{8cm}|}
\hline
\textbf{Direction} & \textbf{Key player} & \textbf{Effects} \\
\hline
Horizontal &  Ionized gas jet & 1. Avoid GC from touching and coalescing with the glycerol in the beaker by bombardment (from Sec. 2.2). \newline 2. Relieve the charge accumulated at both ends of GC which raises the voltage drop across GC and enhances the Kelvin force (Sec. 2.1). \\
\hline
Vertical & Kelvin force & 1. Counteract the weight of GC (Sec. 2.1). \newline 2. Initiate the vertical oscillation that lessens the impact frequency of IGJ and consequently lowers the probability for GC to be torn apart (Sec. 2.2 \& 2.3). \\
\hline
\end{tabular}
\caption{\textbf{Key players and their effects along the horizontal and vertical directions in GC dynamics.}}
\label{tab:GC_effects}
\end{table}

Building on the above findings, GC displays dynamically adaptive behavior under varying IGJ and electric field. Reversible deformation and directional migration were observed across multiple trials (see Supplementary Video 7). In addition, GC experienced the effect of gradient-dependent electric field in KF, with spatial redistribution of forces and internal stress observed under non-uniform fields. These responses were repeatable and varied with both field intensity and stimulation history, indicating an environment-sensitive modulation of GC behavior\cite{kang2024dynamically, zhang2021liquid}.

\subsection{Potential application of GC setup in the railgun}

According the information in previous sections, the deformation of GC in its liquid state leads to asymmetry of the frequency of plasma bombardment in the left and right sides. This aggravates the horizontal movement of GC and prompt its demise by coalescing with the glycerol in the beaker. Therefore, a substitute by a solid object with a relative permittivity greater than 100 can allow us to construct a novel type of railguns. Unlike the conventional electromagnetic railgun \cite{mcnab2005development,xie2021research}, this new system boasts of two advantages: (1) It relies solely on the presence of a strong electric field which is much easier to generate than the magnetic field which requires a powerful current that inevitably incurs large waste of energy via heat. (2) The projectile is kept at the central axis by the plasma bombardment and thus not touching the rail which again saves energy lost via frictions. 

To evaluate the feasibility of this concept, we conduct the same numerical simulations as in Sec. 2.1 by incorporating the air drag. We model the motion of a projectile passing through multiple acceleration units, as illustrated in Fig. 9a. In addition to be responsible for accelerating and firing, the Kelvin force also helps the continual IGJ at keeping the object at the central axis between the electrodes. 

Figure 9b shows that a faster velocity for the projectile can be achieved by adopting a higher permittivity. In the meantime, although the accelerating units are identical, their contribution to the accelerating process decreases as the projectile speeds up and experiences more air drag, as shown in Fig.  9c and d.

\begin{figure}
\centering
\includegraphics[width=8.5cm]{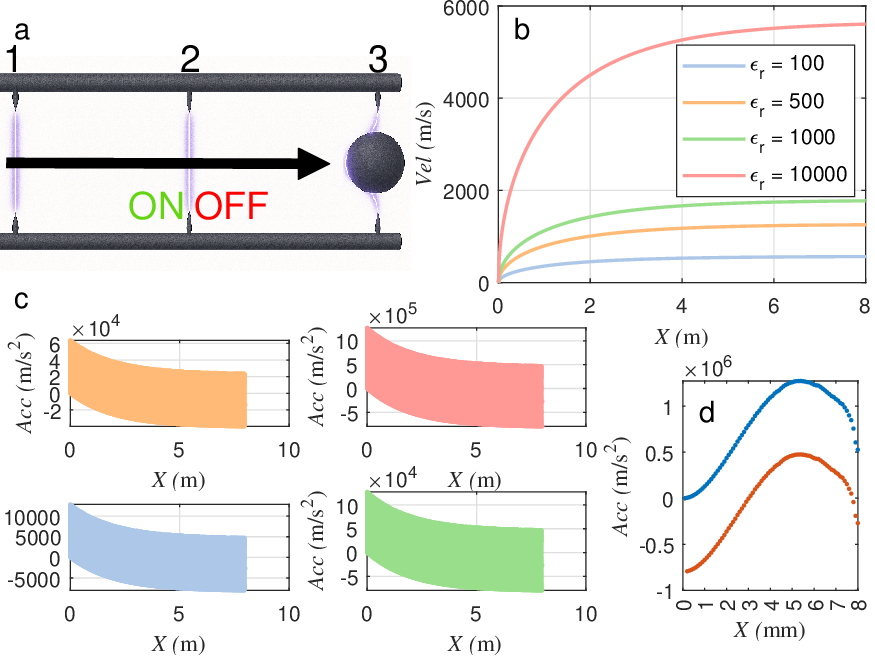}\caption{\textbf{Railgun simulation.} \textbf{a} presents a schematic diagram of railgun composed of three acceleration units each of which provides the Kelvin force via a pair of electrodes. An optical detector can be placed at the midpoint of each electrode pair as a switch to prevent deceleration. \textbf{b.} Simulation results for the velocity of projectile as a function of distance for a railgun consisting of 10,000 accelerating units of voltage 50 kV and spacing 8 mm. Four different relative permittivity values are adopted for the projectile of mass density 10,000 kg/m$^{3}$. \textbf{c.}  shows their acceleration experienced at various positions within the railgun. Note that the actual curve oscillates with a period about 10 mm, as in \textbf{d.} The blue and red curves represent the acceleration due to the first and final units. The reason why the latter is smaller is due to the stronger air drag the projectile experiences after it has been accelerated to a higher speed. 
}
\end{figure}  

\section{Discussion}

Throughout this study, simulations and theoretical models are employed to complement our experiments. 
Section 2.1 confirms that the floating GC is made possible by the interaction of dipoles and the non-uniform electric field, namely, the Kelvin force. It is important to note that surface tension, a mere internal force, does not play any part since the GC is totally detached from the glycerol in both beakers. This casts doubt on the role of surface tension for WB, which according to previous numerical simulations \cite{montazeri2013experimental} contributes equally as the Maxwell tensions, which can lead to the Kelvin force, at countering the weight resistance. 

Unlike most examples involving Taylor cones and droplet deformation, GC exhibits a coupling mechanism that prevents it from being torn apart by the breakup mode  \cite{abbasi2019mono,luo2018breakup}. Analogous to the fuse in electrical circuits, the IGJs act as a safety device, mitigating the stretching force as GC develops under the pull of the electrode, as in Sec. 2.2. In Sec. 2.3, we further showed that the vertical oscillation can alleviates the charge imbalance on both sides of GC in the $X$-direction  and effectively prevent the rupture caused by uneven frequency of plasma bombardment. This is demonstrated by establishing a link between a longer lifetime and a strong correlation for $Y$ and the geometric parameter $r$ and $d$ of GC. Therefore, we strongly recommend using solids that are less prone to deformation and have higher relative permittivity for future experiments, especially for the novel railgun design in section 2.4.

We also observed that the plasma intermittently and randomly bombards the glycerol surface both on GC  and in the beaker, as evidenced by the photos in Fig. 4 \cite{kong2012plasma,xian2009plasma,mericam2009experimental}. These are characteristic signs of IGJ impacts. Previous literature has revealed that weakly ionized plasma jets can significantly stabilize surface indentations on liquids \cite{park2021stabilization}. This explains why, in addition to the centripetal cohesion force resulting from the Kelvin force, GC  structure is not dispersed but rather further stabilized after plasma bombardment. Our setup is analogous to devices that directly apply high voltage to water and glycerol for hydrogen production\cite{zhu2009plasma,choi2012discharge}. Therefore, we reasonably suspect that the plasma discharges on both sides of GC  are also contributing to hydrogen generation.


We did not observe the equivalent of GC in water. What happened was that the water column immediately broke into several droplets when WB became detached to both beakers due to thermal agitations. Since the surface tension coefficient of water is roughly equal to that of glycerol, we suspect this prominence of Rayleigh-Plateau instability in water column is ascribed to its being four orders less viscous. 

 As a potential application of our conclusions, numerical simulations are conducted to demonstrate the feasibility of a new design of railgun. However, existing ceramic materials with relative permittivity greater than 100 have mostly been validated only at the nanometer to micrometer scale. Therefore, the behavior of permittivity decays at the millimeter scale remains unknown. Nonetheless, based on our simulation results in Section 2.4, once the streamlined shape design and the relative permittivity of the accelerated projectile, as indicated in Fig. 9b, are determined, we are confident that our design of mass driver is feasible, cheaper, and simpler. 
 
 We recommend the use of Graph Neural Networks (GNNs)\cite{sanchez2020learning, liu2020learning} to simulate the complex motion of GC in future studies, particularly for modeling the trajectory cannon designs illustrated in Fig. 9\cite{zheng2023deep}. By clarifying the role and interaction among the various factors in the setup, we can modify and find the best design to optimize the lifetime of GC. We have identified the horizontal deformation of GC as the main cause of its instability. Therefore, as part of future work, we will consider replacing GC by a solid material with a high relative permittivity, say above 100 at scales larger than centimeters, which is achievable through the use of ceramic-based composite designs \cite{muhammed2020physical}. The fact that we are operating under a nearly DC source, as opposed to an AC one, reassures us that the induced dipoles can always keep up with the change of electric field.



Inspired by biological systems that adaptively modulate their  properties \cite{yang2021beyond} in response to environmental changes, our glycerol-based construct exhibits dynamic shape transformation under varying plasma stimuli. This behavior aligns with the concept of dynamically adaptive materials as described by Kang (2023)\cite{kang2024dynamically} and liquid-based adaptive materials by Zhang (2021) \cite{zhang2021liquid}, where synthetic matters actively adjust their properties\cite{costa2023smart} rather than remain static under external perturbations. In addition to shape deformation, GC also responds to spatial variations in electric field gradients, where position-dependent modulation of KF contributes to directional movement and asymmetric deformation under non-uniform plasma conditions. 
GC can also provide a new perspective for exploring mean curvature flow \cite{m2017approximations} in the context of rupture evolution, offering potential value for both scientific investigation and educational purposes.


Finally, by encompassing all aspects of droplet coalescence, splitting, floating, and other EHD phenomena, the floating GC opens new avenues for further exploration in the field of EHD, scientific education, and trapping methods, especially as the first occurrence of direct interaction between single droplet deformation and IGJs. It also offers a novel perspective for future developments in electromagnetic-based propulsion mechanisms.

\section{Method}

\subsection{Experiment}
To produce GC under the specified conditions, please follow the procedures below:\\
1. Ensure that the gap between the two beakers  is set at the specified distance. \\
2. Fill the beakers  with glycerol to the brim. \\
3. Apply the specified voltage. \\
4. If there exists a glycerol bridge, please wait patiently until the it heats up and becomes disconnected with the beakers. \\
5. Once the bridge is broken, there is a chance that GC will be formed. Continuously track the processes using the high-speed camera. \\

\subsection{High-Speed Photography}
Data were captured by a high-speed camera  of 3000 fps (Phantom Miro-ex4) and analyzed by Open CV  \cite{culjak2012brief} to track the trajectory of more than 14 sets of GCs in order to find the patterns of their motion under six different parameter sets  chosen between a gap between beakers of 0.6, 0.7 and 0.8 cm and a DC voltage of $V$=12000 and 13000 V. Note that these two voltages, generated by a power supply (HV350REG), are empirically tested to be the most ideal for creating GC. Reliable results are obtained by excluding the first and last 10$\%$ of the recorded data in time.

\subsection{Open CV}
For each video frame of GC, the following pre-processing steps are required to obtain 11 features: \\
1. Apply a mask to select the Region of Interest (ROI) and reduce the influence of other objects, such as the beaker.\\
2. Convert the image to gray scale.\\
3. Apply a 5$\times$5 Gaussian filter kernel.\\
4. Perform edge detection using the Canny operator.\\
5. Smooth the contours by using dilation with one iteration.\\
6. Refine the contours by using erosion with one iteration.\\
7. Utilize the skimage.measure package for further processing.\\
8. Adjust the scale.\\
9. Extract the features.\\

\subsection{Pearson Correlation
Coefficient}
Pearson Correlation
Coefficient (PCCs) provide insights into the interdependence of various GC characteristics over time, and assess the relationship between two variables, quantifying the degree and direction of their association. Denoted by r: 
 \begin{equation}
r = \frac{\sum((X - \bar{X})(Y - \bar{Y}))}{\sqrt{\sum(X - \bar{X})^2} \sqrt{\sum(Y - \bar{Y})^2}}
 \end{equation}
and ranging from -1 to +1, a positive $r$ signifies a positive correlation, meaning both variables tend to change together. Conversely, a negative "$r$" indicates a negative correlation, where one variable tends to increase as the other decreases.


\subsection{Finite Element Method.}
In the numerical simulation, we utilized COMSOL's Electrostatics module for 2D simulations. In Fig. 2A, GC is geometrically approximated by three typical shapes, as depicted in Figs. 2D and 3. The overall thickness in the direction out of this page is set to 0.634 mm, surrounded by air, and an Infinite Element Domain is defined in the outermost domain to simulate an infinitely extending space. The relative permittivity for GC is 42.5, while that for air is 1.

Within the simulation environment, considering the discharge method employed in the experiment, we initially placed electrodes on both side boundaries of GC at $Y$=0, with a spacing of $L$=8 mm between the two point electrodes. We applied a potential difference of 12000V for simulation with the left electrode grounded and the right one at 12000 V. Subsequently, to account for the discharge mode of the plasma, we repositioned the electrodes directly on the surface of GC, corresponding to the original electrode positions. We adjusted their positions along GC's  surface using parametric equations to follow the vertical movement of GC.

For calculating the RT, we obtained the line density value of the Kelvin force in GC's region through Eq. (5), and similarly obtained the line density value of weight, which is 12600 N/m$^{3}$, by integrating over the area within the same region. We then divided the two values to calculate the RT. 

\subsection{Dissipative Particle Dynamics.}

This study employs many-body dissipative particle dynamics (MDPD), a coarse-grained mesoscale simulation technique  \cite{bradski2000dr}. In MDPD, each bead represents a molecular cluster with mass $m$, and particle motion follows Newton's equations, similar to molecular dynamics  \cite{zhao2021review}. There are three types of interaction forces between two beads: conservative force ($f_{ij}^C$), dissipative force ($f_{ij}^D$), and random force ($f_{ij}^R$). The total force exerted on a bead is given by:

\begin{equation}
    \mathbf{F}_i = \sum_{j \neq i} \left( f_{ij}^C + f_{ij}^D + f_{ij}^R \right).
\end{equation}

The conservative force $f_{ij}^C$ in MDPD consists of pairwise attraction and repulsion terms dependent on local density, expressed as:

\begin{equation}
    f_{ij}^C = a_{ij} \omega_c(r_{ij}) \hat{r}_{ij} + b_{ij} (\rho_i + \rho_j) \omega_d(r_{ij}) \hat{r}_{ij},
\end{equation}

where $a_{ij} < 0$ and $b_{ij} > 0$ represent the strength of attraction and repulsion between beads $i$ and $j$, respectively. Here, $r_{ij}$ denotes the inter-bead distance, and $\hat{r}_{ij}$ represents the unit vector connecting beads $i$ and $j$.

The weight functions $\omega_c(r_{ij})$ and $\omega_d(r_{ij})$ decay linearly with the interparticle distance and vanish as $r_{ij} \geq r_c$ and $r_{ij} \geq r_d$, respectively. They are given by:

\begin{equation}
\omega_c(r_{ij}) = 1 - \frac{r_{ij}}{r_c},
\end{equation}

\begin{equation}
\omega_d(r_{ij}) = 1 - \frac{r_{ij}}{r_d}.
\end{equation}

The interaction ranges are simply set as $r_c = 1$ and normally $r_d = 0.75 r_c$. The repulsion is contingent on the local density $\rho_i$ and not pairwise. The local density of bead $i$, denoted as $\rho_i$, is specified as:

\begin{equation}
\rho_i = \frac{15}{2\pi r_d^3} \sum_{j \neq i} \left( 1 - \frac{r_{ij}}{r_d} \right)^2, \quad r_{ij} < r_d,
\end{equation}

\begin{equation}
\rho_i = 0, \quad  r_{ij} > r_d.
\end{equation}

The dissipative force $f_{ij}^D$ and random force $f_{ij}^R$ are defined as in classical dissipative particle dynamics.

The simulations are conducted using reduced units, where all quantities are expressed in non-dimensional form based on the cutoff distance \(r_c\), bead mass \(m\), and thermal energy \(k_B T\). For example, time \(t\) is scaled by

\[
t = \sqrt{\frac{m r_c^2}{k_B T}},
\]

and velocity by

\[
v = \sqrt{\frac{k_B T}{m}}.
\]

The Coulomb force is included in the system, with long-range interactions computed using the Ewald summation technique. All MDPD simulations are performed using the open-source software Large-scale Atomic/Molecular Massively Parallel Simulator (LAMMPS).

The simulation system includes three types of beads: charged electrodes, neutral GC beads, and charged GC beads. The positive and negative electrodes are each represented by a single bead, positioned 40 units apart, and charged \(\pm 2500\) to simulate a strong electric field. The system contains a total of 5488 GC beads, with positive and negative charged GC beads set to 100, 200, 300, 400, or 500, each carrying a charge of \(\pm 0.3\). The charged GC beads are evenly distributed (1:1) between positive and negative charges and positioned closest to the electrodes in the initial configuration. For varying initial charge ratios, the positive charge is fixed, while the negative charge is reduced by 20\%, 40\%, or 60\%.

The initial configurations underwent an equilibration process of 1000 steps, followed by simulations in the canonical (NVT) ensemble. To evaluate various parameters, including the number of charged GC beads, discharge length, and discharging steps, each simulation was run for 2000 cycles. Python code (4.8 section) was integrated into each simulation cycle to randomly discharge or recharge the charged beads, allowing dynamic adjustments that mimic real-world conditions. As a result, the interactions between GC beads and the strong electrode behavior were thoroughly investigated.

\subsection{Simulation of IGJs}

For each chargeable bead (charged GC beads), perform the following process to simulate plasma: Define the plasma bombardment probability $P(x)=(x/L)^{2n}$, where $x$ is the x-coordinate of the bead, $L$ is a given length scale, and $n$ is set to 25. At every $k$-th timestep, for each chargeable bead $i$, generate a random number $\kappa$ uniformly distributed between 0 and 1. The bead is then assigned a charge of $-sgn(x_i)q\Theta(\kappa-P(x_i))$, where $sgn(x)$ is the sign function, and $\Theta(x)$ is the Heaviside step function.

\subsection{Numerical Algorithm for Simulating Railgun}

To evaluate the feasibility of using polarized materials as a projectile under a strong nonuniform electric field, we developed a numerical algorithm that simulates the velocity and acceleration profiles as the projectile traverses multiple plasma-based acceleration units. The algorithm incorporates air drag and uses the RT as a function of thrust, derived from simulation data.

An ellipsoidal particle with major axis $a = 2.268 \times 10^{-3}$ m and minor axis $b = 0.634 \times 10^{-3}$ m was assumed, corresponding to the observed GC geometry. The mass was calculated from the volume of the ellipsoid and an estimated mass density of $10^4$ kg/m. The air drag was computed with a drag coefficient $C_d = 0.35$ and air density $\rho = 1.225$ kg/m$^3$.

The simulation proceeds through 10,000 iterations to ensure convergence of terminal velocity. At each iteration, Euler integration was employed over a spatial mesh with resolution $\Delta x$, where the velocity $v_{i+1}$ is updated by:

\[
v_{i+1} = v_i + a_i \cdot \frac{\Delta x}{v_i},
\]
and the instantaneous acceleration $a_i$ is calculated via Newton's second law:

\[
a_i = \frac{F_{\text{thrust}} - F_{\text{drag}}}{m},
\]
where $F_{\text{thrust}} = RT \cdot m \cdot g$ and $F_{\text{drag}} = \frac{1}{2} \rho C_d A v_i^2$. The process iterates until the terminal velocity stabilizes.

\section{Author contributions}
Experiment: Ping-Rui Tsai, Hong-Yue Huang, Yu-Ting Cheng, Cheng-Wei Lai, Yu Hsuan Kao.\\
Dissipative Particle Dynamics.: Yu-Jane Sheng, Chih-Jung Lin, Bo-Kai, Xu\\
Open CV: Hong-Yue Huang, Jih-Kang Hsieh, Yu Hsuan Kao.\\
Figure: Yu Hsuan Kao.\\
Theory: Ping-Rui Tsai, Tzay-Ming Hong.\\
Writing: Ping-Rui Tsai, Tzay-Ming Hong.\\
COMSOL Finite Element Method (FEM) simulation efforts: Ying-Pin Tsai, Wen-Chi Chen, Fu-Li Hsiao.\\
Technical assistance: Po-Heng Lin.\\

\section{Acknowledgements}
We would like to thank Prof. Keh-Chyang Leou for useful knowledge on plasma, and Prof. Jow-Tsong Shy for technical supports. Financial support from the National Science and Technology Council in Taiwan under Grant No. 111-2112-M007-025 and No. 112-2112-M007-015 is also gratefully acknowledged.

\section{Conﬂict of Interest}
The authors declare no conﬂict of interest.

\section{Supplementary information}
In this letter, we have one SIGuide.doc and seven Supplementary Videos. 

\section{Correspondence and requests for materials}
 Should be addressed to Ping-Rui Tsai or Tzay-Ming Hong.

\bibliography{sn-bibliography}

\end{document}